\documentstyle[aps,prl,preprint,floats,epsfig]{revtex}

\begin{document}

\preprint{\tighten\vbox{\hbox{\hfil CLNS 01/1739}
                        \hbox{\hfil CLEO 01-11}
}}

\title{\large Improved Upper Limits on the FCNC Decays
$B \rightarrow K \ell^+ \ell^-$ and $B \rightarrow K^*(892) \ell^+ \ell^-$}

\author{(CLEO Collaboration)}

\date{ 14 June 2001}
\maketitle
\tighten

\begin{abstract}
We have searched a sample of 9.6 million $B \bar B$ events for the
flavor-changing neutral current decays $B \rightarrow K \ell^+ \ell^-$ and
$B \rightarrow K^*(892) \ell^+ \ell^-$.  We subject the latter decay to the
requirement that the dilepton mass $m_{\ell \ell}$ exceed 0.5 GeV.  There is no
indication of a signal.  We obtain the 90\% confidence level upper limits
${\cal B}(B \rightarrow K \ell^+ \ell^-) < 1.7 \times 10^{-6}$ and
${\cal B}(B \rightarrow K^*(892) \ell^+ \ell^-)_{m_{\ell \ell} > 0.5 {\rm GeV}}
 < 3.3 \times 10^{-6}$.  We also obtain an upper limit on the weighted average
$0.65 {\cal B}(B \rightarrow K \ell^+ \ell^-) + 0.35
{\cal B}(B \rightarrow K^*(892) \ell^+ \ell^-)_{m_{\ell \ell} > 0.5 {\rm GeV}}
 < 1.5 \times 10^{-6}$.  The weighted-average limit is only 50\% above the
Standard Model prediction.
\end{abstract}
\newpage
{
\renewcommand{\thefootnote}{\fnsymbol{footnote}}
\begin{center}
S.~Anderson,$^{1}$ V.~V.~Frolov,$^{1}$ Y.~Kubota,$^{1}$
S.~J.~Lee,$^{1}$ R.~Poling,$^{1}$ A.~Smith,$^{1}$
C.~J.~Stepaniak,$^{1}$ J.~Urheim,$^{1}$
S.~Ahmed,$^{2}$ M.~S.~Alam,$^{2}$
% ...
\end{center}
 
\small
\begin{center}
$^{1}${University of Minnesota, Minneapolis, Minnesota 55455}\\
$^{2}${State University of New York at Albany, Albany, New York 12222}\\
% ...
$^{13}${Virginia Polytechnic Institute and State University,
Blacksburg, Virginia 24061}\\
% ...
$^{25}${University of Kansas, Lawrence, Kansas 66045}
\end{center}
\setcounter{footnote}{0}
\newpage 


    The flavor-changing neutral current (FCNC) decay
$b \rightarrow s \ell^+ \ell^-$ is
sensitive to physics beyond the Standard Model\cite{bsm-theory}, and, like the
radiative penguin decay $b \rightarrow s \gamma$, is more amenable to
calculation than
purely hadronic FCNC decays.  The decay $b \rightarrow s \ell^+ \ell^-$  depends
on the magnitude and sign of the three Wilson coefficients $C_7$, $C_9$, and
$C_{10}$ in the effective Hamiltonian, while $b \rightarrow s \gamma$ depends
only on the magnitude of $C_7$.  These three Wilson coefficients are likely
places for New Physics to appear, as they come from loop and box diagrams.
Upper limits on the branching fraction for
$b \rightarrow s \ell^+ \ell^-$ thus place constraints on New Physics, while
observation of $b \rightarrow s \ell^+ \ell^-$ at a rate in excess of that
predicted by the Standard Model would provide evidence for New Physics.
Just as we first observed $b \rightarrow s \gamma$
through its exclusive decay $B \rightarrow K^*(892) \gamma$\cite{Kstargamma},
and only later in an inclusive fashion\cite{bsgamma-orig}, so we search
for $b \rightarrow s \ell^+ \ell^-$ through the decays
$B \rightarrow K \ell^+ \ell^-$ and $B \rightarrow K^*(892) \ell^+ \ell^-$.
Existing published upper limits come from CDF\cite{CDF}, and from earlier work
by us\cite{CLEO-old}.  Here we present improved upper limits.

    The $B \rightarrow K^* \ell^+ \ell^-$ decay rate\footnote{Throughout this
Letter, the symbol $K^*$ means $K^*(892)$.} peaks at low dilepton mass
$m_{\ell \ell}$, due to the photon pole from
$B \rightarrow K^* \gamma_{virtual},\ 
\gamma_{virtual} \rightarrow \ell^+ \ell^-$.  Because the decay
$B \rightarrow K^* \gamma$ is already well studied, and $\vert C_7 \vert$
thus reasonably well known, we require $B \rightarrow K^* \ell^+ \ell^-$
candidates to have dilepton mass above 0.5 GeV, to reduce the contribution
from the virtual photon diagram and thus the dependence on $\vert C_7 \vert$.

    The data used in this analysis were taken with the CLEO
detector\cite{detector} at the Cornell Electron Storage Ring (CESR), a symmetric
$e^+ e^-$ collider operating in the $\Upsilon({\rm 4S})$ resonance region. The
data sample consists of 9.2 ${\rm fb}^{-1}$ at the resonance, corresponding to
9.6 million $B \bar B$ events,
and 4.5 ${\rm fb}^{-1}$ at a center-of-mass energy 60 MeV below the resonance.
The sample below the resonance provides information on the background from
continuum processes $e^+ e^- \rightarrow q \bar q,\ q = u,d,s,c$.

    We search for $B \rightarrow K^{(*)} \ell^+ \ell^-$ both in the
$\mu^+ \mu^-$ and $e^+ e^-$ modes, for $B \rightarrow K \ell^+ \ell^-$ in both
the $K^\pm$ and $K^0$ modes, and for $B \rightarrow K^* \ell^+ \ell^-$ in
the $K^{*0} \rightarrow K^+ \pi^-$ and $K^0 \pi^0$ modes and in
the $K^{*\pm} \rightarrow K^\pm \pi^0$ and $K^0 \pi^\pm$ modes, a total of 12
distinct final states.  $K^0$ is detected via the
$K^0 \rightarrow K^0_S \rightarrow \pi^+ \pi^-$ decay chain.

    There are three main sources of background:
$B \rightarrow K^{(*)} \psi^{(\prime)},\ \psi^{(\prime)} \rightarrow \ell^+
\ell^-$, and other $B \rightarrow \psi^{(\prime)} X$ decays; continuum processes
with two apparent leptons (either real leptons or hadrons misidentified as
leptons); and
$B \bar B$ decays other than $B \rightarrow \psi^{(\prime)} X$, with two
apparent leptons.  We suppress $B \rightarrow K^{(*)} \psi^{(\prime)}$
events with vetoes on the dilepton mass around $\psi$ and $\psi^\prime$, in
particular $2.80 < m_{ee} < 3.23$ GeV,
$2.90 < m_{\mu \mu} < 3.20$ GeV, $3.51 < m_{ee} < 3.77$ GeV, and
$3.55 < m_{\mu \mu} < 3.74$ GeV.  These very wide cuts are needed because of the
low-side radiative tail from internal and external bremsstrahlung from
$\psi^{(\prime)} \rightarrow e^+ e^-$, and to a lesser extent the low-side
radiative tail from internal bremsstrahlung from the $\mu^+ \mu^-$ decay.

    We suppress the background from $B \bar B$ semileptonic decays with a cut
on the event missing energy, $E_{miss}$, since events with leptons from
semileptonic $B$
or $D$ decay  contain neutrinos.  Distributions in $E_{miss}$ for Monte
Carlo samples of signal and background are shown in Fig.~\ref{fig:Emiss}.
We suppress continuum events with a cut on a Fisher discriminant, a linear
combination of $R_2$ (the ratio of second and zeroth Fox-Wolfram
moments\cite{Fox-Wolfram} of the
event), $\cos \theta_{tt}$ ($\theta_{tt}$ the angle between the
thrust axis of the candidate $B$ and the thrust axis of the rest of the event),
S (the sphericity), and $\cos \theta_B$ ($\theta_B$ the production angle of the
candidate $B$, relative to the beam direction).  In particular,
${\cal F} = R_2 + 0.117 \vert \cos \theta_{tt} \vert + 0.779 (1 - S)
+ 0.104 \vert \cos \theta_B \vert$.  The coefficients of all terms but $R_2$
were determined by the standard Fisher discriminant procedure\cite{fisher}.
The relative weight given to $R_2$ was determined visually, from a
scatter plot of $R_2$ {\it vs.} the Fisher discriminant from the other three
variables.  Distributions in ${\cal F}$ for Monte Carlo samples of signal and
background are shown in Fig.~\ref{fig:Emiss}.

    For those decay modes involving a charged kaon, we use specific ionization 
($dE/dX$) and time-of-flight information to identify the kaon, cutting loosely
(3 standard deviations) if those variables deviate from the mean for kaons in
the direction away from the mean for pions, and harder (by a variable number,
denoted $kID^{cut}$, of standard deviations) if they deviate on the side towards
the pions.

    All cuts have been determined from Monte Carlo samples: continuum events,
$B \bar B$ events with no signal, and $B \bar B$ events with a
$B \rightarrow K^{(*)} \ell^+ \ell^-$ signal.  We optimized cuts on ${\cal F}$,
$E_{miss}$, and, where appropriate, kaon identification simultaneously.  We
found that the optimum curve in efficiency {\it vs.} background space was traced
out if we required that cuts on ${\cal F}$, $E_{miss}$, and $kID$ were tightened
or loosened together.  In particular, we found that the optimum curve was well
described by $E^{cut}_{miss} = 1.0 + 6.67 \times ({\cal F}^{cut} - 0.8)$ GeV, 
$kID^{cut} = 3.0 - 0.66 \times (3.0 - E^{cut}_{miss})$.  With this formulation,
we have a single cut variable, ${\cal F}^{cut}$, to optimize.

    We optimized the cuts to obtain either the best upper limit, assuming no
signal, or to  see the smallest possible signal.  These two different
optimization  procedures led to similar cuts, and we took the average.  In the
optimization we allowed the value of ${\cal F}^{cut}$ to vary from decay mode
to mode.  The final cuts on ${\cal F}$ are shown in Table~\ref{tab:limits}.
The corresponding cuts on $E_{miss}$ and $kID$ can be obtained from the
expressions given above.

    Our final discrimination between signal and background comes from the
$B$ reconstruction variables conventionally used for decays from the
$\Upsilon(4S)$, beam-constrained mass
$M_{cand} \equiv \sqrt{E^2_{beam} - P^2_{cand}}$ and
$\Delta E \equiv E_{cand} - E_{beam}$.
Our resolution in $M_{cand}$ is 2.5 MeV, and in $\Delta E$, 20 MeV.  We define a
signal box in $M_{cand} - \Delta E$ space, $\pm$ 6.5 MeV around the $B$ mass
by $\pm$ 60 MeV ($\mu^+ \mu^-$) or $\pm$ 70 MeV ($e^+ e^-$).  The signal box
for $e^+ e^-$ events is shifted off zero by 5 MeV in $\Delta E$, because
radiation losses cause $B \rightarrow K^{(*)} e^+ e^-$ signal events to peak at
--5 MeV in $\Delta E$ rather than at zero.

Background from $B \rightarrow \psi^{(\prime)} X$ is estimated from
Monte Carlo simulation.  Background from other $B$ decay processes and from
continuum processes is determined using a large sideband region in
$M_{cand} - \Delta E$ space:
$5.20 < M_{cand} < 5.29$ GeV, $\vert \Delta E \vert < 0.25$ GeV, but excluding
the signal box.  From Monte Carlo simulation, we found that the ratio of events
in the signal region to events in the sideband region is 0.024 for $B \bar B$
background events other than $B \rightarrow \psi^{(\prime)} X$, and 0.027 for
continuum background, in both cases smaller than the ratio of areas,
0.038, because backgrounds fall off as the candidate mass approaches the beam
energy.  Recognizing that this ratio must be larger for continuum background
than for $B \bar B$ background because the more jet-like continuum events will
not fall off as rapidly as candidate mass approaches beam energy, and
recognizing that using a lower background in an upper limit computation gives a
more conservative answer, we take the continuum scaling factor to be equal to
the $B \bar B$ scaling factor, rather than using 0.027.

    The $e^+ e^-$ events suffer a degradation in resolution due to internal and
external bremsstrahlung from the electrons.  We partially recover that
resolution by adding to each electron energy the energy of those photons found
nearby in angle. This procedure improves the $\psi$ veto, and resolution in
$E_{miss}$, $M_{cand}$, and $\Delta E$.

    The number of events that satisfy all cuts and land in the signal box is
given, for each mode, in Table~\ref{tab:limits},  along with the background
estimate.  We find 3 $B \rightarrow K \ell^+ \ell^-$ candidates, with an
expected background of 2.0; we find 4 $B \rightarrow K^* \ell^+ \ell^-$
candidates, with an expected background of 3.8.  Thus there are a total of 7
events with an expected background of 5.8.  The probability that a true mean of
5.8 will fluctuate up to 7 or more events is 36\%.  Thus, there is no indication
of signal.  A scatter plot of $M_{cand}$ {\it vs.} $\Delta E$ for events passing
all other cuts and landing in the signal or sideband region is shown in
Fig.~\ref{fig:scatter}.  Again, no indication of a signal.  We obtain upper
limits.

    We calculate upper limits, at 90\% confidence level, taking backgrounds into
account\cite{oldPDG}.  To allow for the uncertainty in the background estimate,
we use a value for the background which is reduced below our actual estimate.
For individual modes, we use half the estimated background. For
$B \rightarrow K \ell^+ \ell^-$ and $B \rightarrow K^* \ell^+ \ell^-$ totals,
we use the estimated background
minus 1.28 standard deviations of its combined statistical and systematic error.
(The factor 1.28 gives a 90\% confidence level lower estimate of the background,
assuming its uncertainty has a Gaussian distribution.)
Results are given in Table~\ref{tab:limits}.

    We gain experimental sensitivity to the underlying
$b \rightarrow s \ell^+ \ell^-$ interaction by calculating a weighted average
over the two decay modes studied.  To account for the difference in the
experimental precision for each mode,  we weight them by our relative
efficiencies, that is, we compute an upper limit on the sum over all the
individual sub-modes.  This gives an upper limit on
$0.65 {\cal B}(B \rightarrow K \ell^+ \ell^-) +
0.35 {\cal B}(B \rightarrow K^* \ell^+ \ell^-)$, where the coefficients 0.65
and 0.35 are the relative efficiencies we have for the two modes.

    We use Monte Carlo simulation to determine the efficiency for detecting the
signal modes.  The decays $B \rightarrow K \ell^+ \ell^-$ and
$B \rightarrow K^* \ell^+ \ell^-$ are generated using the model of
Ali {\it et al.}\cite{bsm-theory}.  The helicity of the $K^*$ is taken into
account.  Final-state radiation is included, using the CERNlib subroutine
{\sc Photos}\cite{photos}.  We have also generated decays with the two extreme
variations that Ali {\it et al.} suggest for their model, and with several other
models\cite{otherModels,bsm-theory}.
We find the model-to-model variation in efficiency to 
be small, with a relative r.m.s. variation of $\pm$3\%.

    To check our procedures, we have looked for the decays
$B \rightarrow J/\psi K^{(*)}$ rather than vetoing them.  We compare the 
branching fractions obtained with those from prior
CLEO measurements.  There is good agreement --  differences are at or below the
one-standard-deviation level.

    Systematic errors are of two varieties -- those on the estimate of
signal detection efficiencies, and those on the estimate of backgrounds.  The
contributors to the former are lepton identification uncertainties
(contributing $\pm$5\%, relative,  in the efficiency), missing-energy-simulation
uncertainties ($\pm$3.5\%), and simulation uncertainties for $M_{cand}$,
$\Delta E$, $dE/dX$, time of flight, and ${\cal F}$ ($\pm$3.0\%),
giving a $\pm$7\% relative uncertainty in the overall efficiency.  To this we
add in quadrature $\pm$3\% for the model dependence of the efficiency, discussed
earlier. The contributors to the background uncertainties are the 
modelling of $B \rightarrow \psi^{(\prime)} X$ ($\pm$10\%), and uncertainties
in the scale factors from sideband region to signal region in
$M_{cand}\ -\ \Delta E$ space.  We assign a systematic error to the
$B \bar B$ background scale
factor by determining it with different methods, obtaining 0.024 $\pm$ 0.004.
Recognizing that the scale factor for continuum should be larger than that for
$B \bar B$, we conservatively set it equal to the $B \bar B$
scale factor, with the same (correlated) systematic error.
The errors shown on the backgrounds in Table~\ref{tab:limits}
include statistical errors and the systematic errors just described.

    There is no universally agreed-upon procedure for including systematic
errors in upper-limit estimates.  We conservatively reduce the background by
1.28 standard deviations, and decrease the efficiency by 1.28 standard
deviations.  In this way we obtain our final results:

\newpage

$${\cal B}(B \rightarrow K \ell^+ \ell^-) < 1.7 \times 10^{-6}\ ,$$

$${\cal B}(B \rightarrow K^*(892) \ell^+ \ell^-)_{m_{\ell \ell}>0.5 {\rm GeV}}
< 3.3 \times 10^{-6}\ ,\ \ {\rm and}$$

$$0.65{\cal B}(B \rightarrow K \ell^+ \ell^-) +
0.35{\cal B}(B \rightarrow K^*(892) \ell^+ \ell^-)_{m_{\ell \ell}>0.5 {\rm GeV}}
 < 1.5 \times 10^{-6}\ ,$$
\noindent all at 90\% confidence level.  These results are significant
improvements over previously published limits\cite{CDF,CLEO-old}.

    The Standard Model values for these branching fractions, as given by Ali
{\it et al.}\cite{bsm-theory}, are $0.6 \times 10^{-6}$ for
${\cal B}(B \rightarrow K \ell^+ \ell^-)$ and $1.8 \times 10^{-6}$ for
${\cal B}(B \rightarrow K^* \ell^+ \ell^-)_{m_{\ell \ell}>0.5 GeV}$, and thus
$1.0 \times 10^{-6}$ for the 0.65 / 0.35 weighted average.  The limit on the
branching fraction for $B \rightarrow K \ell^+ \ell^-$ is therefore
about three times its Standard Model prediction, the limit on the branching
fraction for $B\rightarrow K^* \ell^+ \ell^-$, subject to the requirement that
$m_{\ell \ell} > 0.5$ GeV, is about twice its Standard Model prediction, and
the 0.65 / 0.35 weighted average is only 50\% larger than its Standard
Model prediction.

    In summary, we have searched for the decays $B \rightarrow K \ell^+ \ell^-$
and $B \rightarrow K^*(892) \ell^+ \ell^-$.  We find no indication of a signal,
and obtain upper limits on the branching fractions.  These limits are
consistent with Standard Model predictions, but not far above them.

We gratefully acknowledge the effort of the CESR staff in providing us with
excellent luminosity and running conditions.
This work was supported by 
the National Science Foundation,
the U.S. Department of Energy,
the Research Corporation,
the Natural Sciences and Engineering Research Council of Canada,
the Texas Advanced Research Program, and the Basic Science program of the Korea
Research Foundation.

\begin{table}[ht]
\begin{center}
\begin{tabular}{|c|c|c|c|c|c|}
      mode  & $ {\cal F}^{cut}$ & observed &  background & efficiency &
${\cal B} \times 10^6$  \\
        & & events  &   &        & upper Lim. \\ \hline
$K^0 e^+ e^-$         & 0.938 &   1   & 0.10   & 0.053   & 7.6 \\
$K^0 \mu^+ \mu^-$     & 0.925 &   0   & 0.21   & 0.041   & 7.8 \\
$K^+ e^+ e^-$         & 0.925 &   1   & 0.95   & 0.165   & 2.3 \\
$K^+ \mu^+ \mu^-$     & 0.850 &   1   & 0.74   & 0.111   & 3.4 \\ \hline
$K \ell^+ \ell^-$     &       &   3   & 1.99 $\pm$0.35   & 0.370 &1.49 \\ \hline
$K^0 \pi^+ e^+ e^-$   & 0.925 &   0   & 0.35   & 0.019   & 12.8 \\
$K^0 \pi^+ \mu^+ \mu^-$& 0.900 &  0   & 0.27   & 0.015   & 15.6 \\
$K^+ \pi^0 e^+ e^-$   & 0.800 &   3   & 0.27   & 0.015   & 46.0 \\
$K^+ \pi^0 \mu^+ \mu^-$& 0.750 &  0   & 0.49   & 0.008   & 29.3 \\
$K^+ \pi^- e^+ e^-$   & 0.925 &   1   & 0.97   & 0.071   & 5.0 \\
$K^+ \pi^- \mu^+ \mu^-$& 0.875 &  0   & 1.24   & 0.052   & 4.6 \\
$K^0 \pi^0 e^+ e^-$   & 0.900 &   0   & 0.11   & 0.007   & 35.8 \\
$K^0 \pi^0 \mu^+ \mu^-$& 0.750 &  0   & 0.10   & 0.002   & 117.3 \\ \hline
$K^* \ell^+ \ell^-$     &       &   4   & 3.80$\pm$0.57 & 0.188 & 2.94 \\ \hline
Sum                   &       &   7   & 5.79$\pm$0.83   & 0.558 & 1.35 \\
\end{tabular}
\end{center}
\caption{
Value of ${\cal F}^{cut}$, number of events observed in signal window,
the expected background, efficiency, and upper limit on the branching
fraction, for the 12 modes, and for $B \rightarrow K \ell^+ \ell^-$,
$B \rightarrow K^* \ell^+ \ell^-$, and 
0.65${\cal B}(B \rightarrow K \ell^+ \ell^-) + 
0.35{\cal B}(B \rightarrow K^* \ell^+ \ell^-)$.  Note, the upper limits given in
this table have been corrected for systematic error in background, but {\it not}
corrected for the systematic error in efficiency.
\label{tab:limits}}
\end{table}

\begin{figure}
\begin{center}
\epsfysize=6.0in
\epsfbox{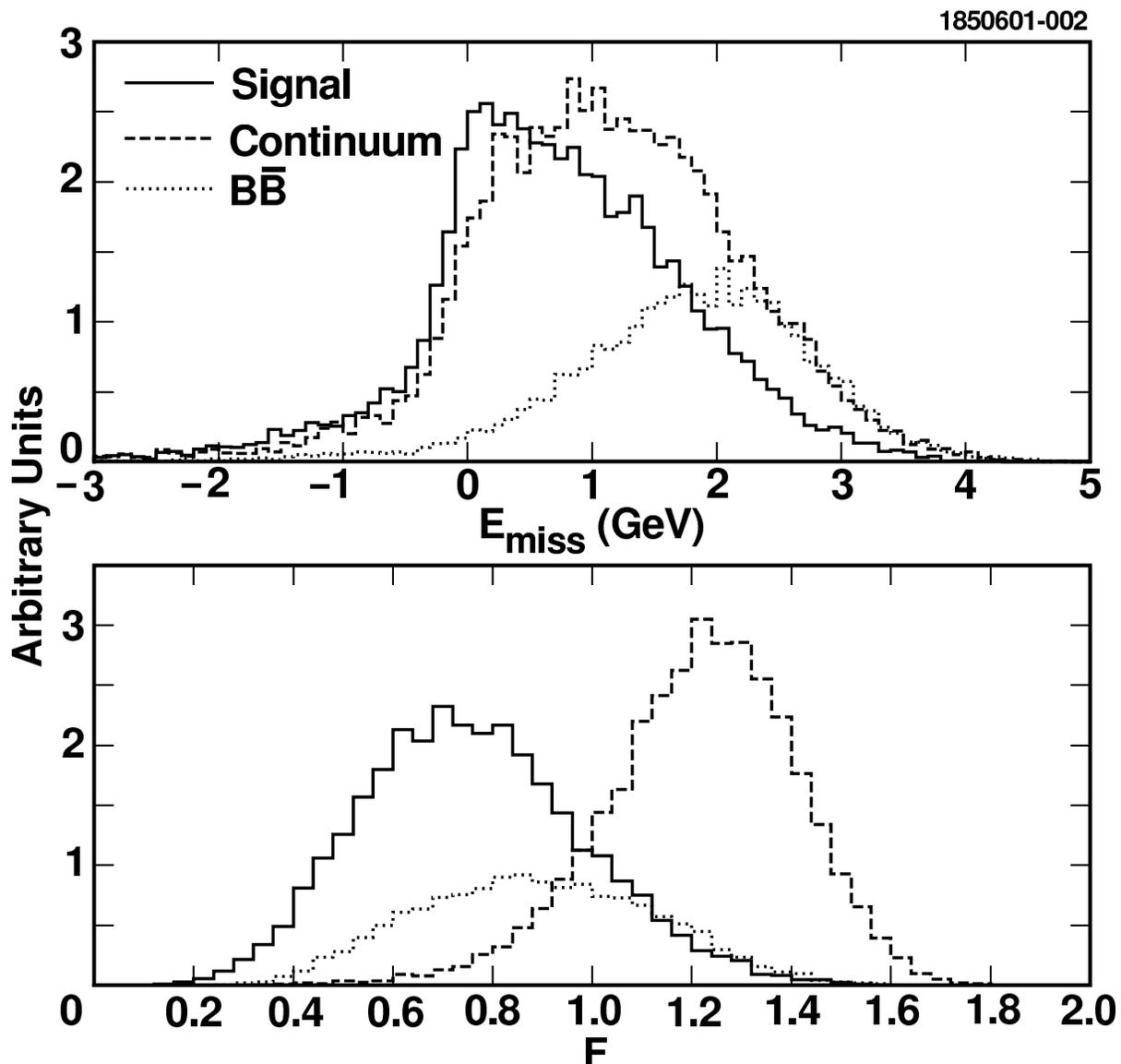}
\hfill \caption{
Distributions in $E_{miss}$ (upper) and ${\cal F}$ (lower) for Monte Carlo
samples of signal events (solid), $B \bar B$ background events (dotted), and
continuum background events (dashed).  The vertical scale is arbitrary.
\label{fig:Emiss}}
\end{center}
\end{figure}

\begin{figure}
\begin{center}
\epsfysize=5.0in
\epsfbox{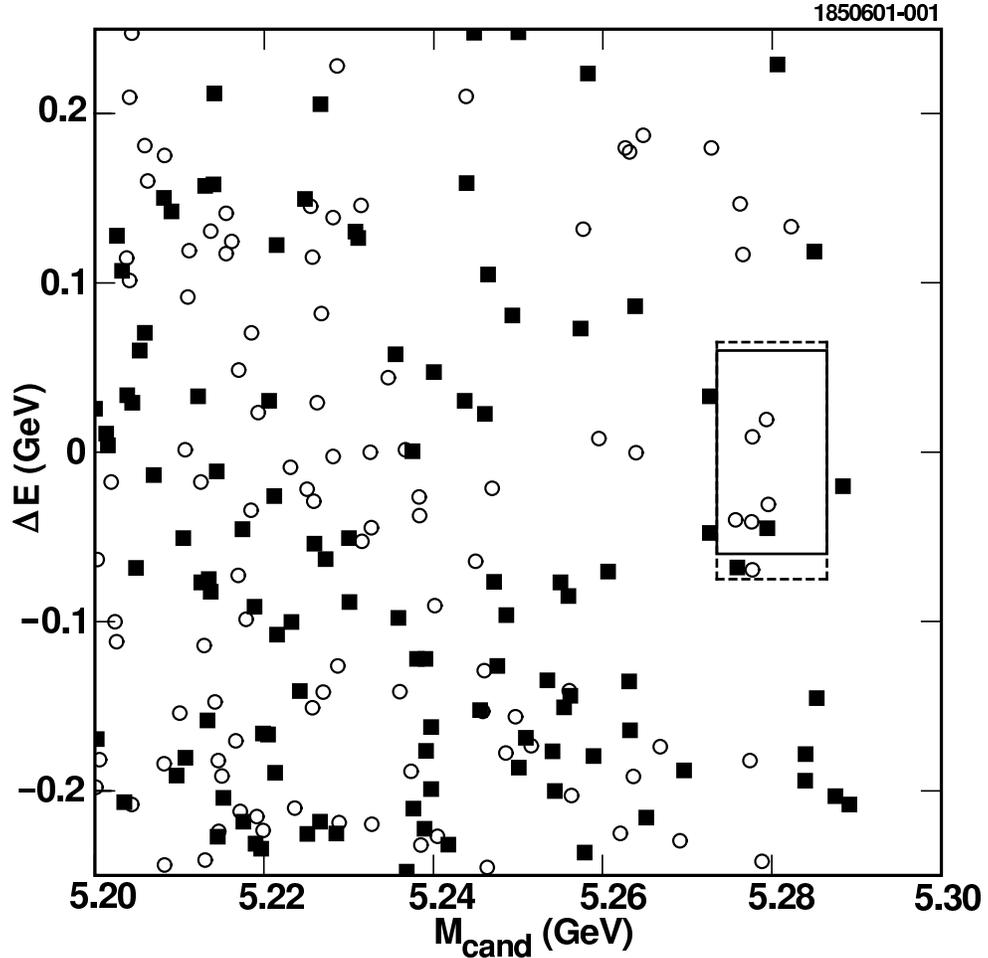}
\hfill \caption{
Scatter plot of $M_{cand}$ {\it vs.} $\Delta E$ for $e e$ events (open circles)
and $\mu \mu$ events (solid squares) passing all other cuts, for data on the
$\Upsilon(4S)$ resonance.  The smaller box (solid) is the $\mu^+ \mu^-$ signal
box, while the larger box (dashes), shifted 5 MeV toward negative $\Delta E$,
is the $e^+ e^-$ signal box.
\label{fig:scatter}}
\end{center}
\end{figure}

\end{document}